\documentclass[12pt]{JHEP}

\usepackage{graphicx}
\usepackage{bm}
\usepackage{amsmath,epsfig}
\usepackage{amssymb,amsfonts}
\usepackage{latexsym}
\usepackage{epsfig}
\def\hri#1#2{\href{http://arxiv.org/abs/#1}{[ArXiv:#1]#2}}
\def\hre#1#2{\href{http://arxiv.org/abs/#1/#2}{[ArXiv:#1/#2]}}

\def\be{\begin{equation}}
\def\ee{\end{equation}}
\def\bea{\begin{eqnarray}}
\def\eea{\end{eqnarray}}

\newcommand\fverb{\setbox\pippobox=\hbox\bgroup\verb}
\newcommand\fverbdo{\egroup\medskip\noindent%
                        \fbox{\unhbox\pippobox}\ }
\newcommand\fverbit{\egroup\item[\fbox{\unhbox\pippobox}]}

\newcommand{\la}{\lambda}

\newcommand{\bear}{\begin{eqnarray}}
\newcommand{\eear}{\end{eqnarray}}

\newbox\pippobox

\def\ie{{\it i.e.~}}
\def\lab{\label}
\def\6{\partial}

\def\a{\alpha}
\def\nn{\nonumber}

\def\le{\left}
\def\ri{\right}

\def\m{\mu}
\def\n{\nu}

\def\sq
\def\a{\alpha}

\def\l{\lambda}
\def\La{\Lambda}

\def\tr{{\rm Tr}}

\def\eps{\epsilon}

\def\la{\langle}
\def\ra{\rangle}

\def\ka{\kappa}

\title{Deconfinement and Gluon
Plasma Dynamics in Improved Holographic QCD}

\author{U. G{\"u}rsoy$^{1,2}$,
E. Kiritsis$^{1,3}$, L. Mazzanti$^1$, F. Nitti$^1$\\
$^1$CPHT, Ecole Polytechnique, CNRS,
 91128, Palaiseau, France\\
 ( UMR du CNRS 7644).\\
~\\
$^2$Laboratoire de Physique Th\'eorique,\\
Ecole Normale Sup\'erieure,
24, Rue Lhomond, Paris 75005, France.\\
~\\
$^3$Department of Physics, University of Crete\\
71003 Heraklion, Greece\\
~\\}

\preprint{0804.0899 [hep-th]\\CPHT-RR024.0408}      

\abstract{The finite temperature physics of the pure glue sector
in the improved holographic QCD model of
\href{http://arxiv.org/abs/0707.1324}{ArXiv:0707.1324}  and   \href{http://arxiv.org/abs/0707.1349}{ArXiv:0707.1349} is addressed.
 The thermodynamics of 5D dilaton gravity duals to confining
gauge theories is analyzed.
  We show that they exhibit a first order Hawking-Page type
phase transition. In the explicit background
  of  \href{http://arxiv.org/abs/0707.1349}{ArXiv:0707.1349},
we find $T_c = 235$ MeV. The temperature dependence of various thermodynamic quantities such as the pressure, entropy and  speed of sound is calculated.
The results show a good agreement with the corresponding lattice data.}
\begin{document}

\maketitle

\section{Introduction}

Despite several decades' efforts, an important part of the dynamics of QCD
remains far from analytical control and in several cases numerical techniques have proved too difficult to implement.
In particular, recent experiments at RHIC seem to probe dynamical properties of the Quark Gluon Plasma (QGP)
phase which are not within the reach of lattice techniques without extra assumptions.

On the other hand large-$N_c$ techniques have promised early-on an alternative approach to the strongly coupled physics of QCD
based on an effective string theory description of glue.
This route took an interesting twist in 1997 with the advent of the Maldacena conjecture \cite{malda}, with the unexpected
result that the string theory must live in more than four dimensions. In particular there is one extra dimension, known as the holographic
dimension, that plays the role of (renormalization group) energy scale of the strongly coupled gauge theory.

Since \cite{malda}
 there has been a flurry of attempts to devise such correspondences for gauge theories with less supersymmetry with the obvious final goal: QCD.
 Several interesting string duals with a QCD-like low lying spectrum and confining IR physics were proposed \cite{D4}.
In the simplest $D4$ example flavor can be added via the addition of $D_8$ branes \cite{sas} and its finite temperature phase structure
has similarities with QCD \cite{weiz}.
Although such theories reproduced the qualitative features of IR QCD dynamics,
they contain Kaluza-Klein  modes, not expected in QCD,
with KK masses of the same order as the dynamical scale
of the gauge theory. Above this scale the theories deviate from QCD.
Despite the hostile environment  of  non-critical theory,
several attempts have been made  to
understand holographic physics in lower dimensions
 in order to avoid the KK contamination,
based on two-derivative gravitational actions, \cite{ks}.

 A different and more phenomenological approach was in the meantime developed and is now known as AdS/QCD.
The original idea described  in \cite{ps} was successfully applied to the meson
sector in \cite{adsqcd1}, and its thermodynamics was analyzed in \cite{herzog}.
The bulk gravitational background consists of a slice of AdS$_5$, and a constant dilaton.
There is a UV and an IR cutoff. The confining IR physics is imposed by boundary
conditions at the IR boundary.
This approach, although crude,  has
been partly successful in studying meson
physics, despite the fact that the dynamics driving chiral symmetry breaking must be
imposed by hand via IR boundary conditions.
Its shortcomings however include a glueball spectrum that does not fit well the lattice data,
the fact that magnetic quarks are confined instead of screened, and asymptotic Regge trajectories for
glueballs and mesons are quadratic instead of linear. A phenomenological fix of the last problem was suggested by
introducing a soft IR wall, \cite{soft}. Although this fixes the asymptotic spectrum, it does not allow a proper
treatment of thermodynamics. In particular, neither dilaton nor metric  equations of motion are solved.
Therefore the ``on-shell" action is not really on-shell. The entropy computed from the BH horizon does
not match the entropy calculated using standard thermodynamics from the
free energy computed from the action, etc.
Phenomenological metrics for the deconfined phase were also suggested, \cite{adr,keijo}
capturing some aspects of the expected thermodynamics.

In \cite{ihqcd} an improved model for QCD was proposed. It united inputs from both gauge theory and string theory
while keeping the simplicity of a two derivative action. It could describe both the region of asymptotic freedom as well as
the strong IR dynamics of QCD.  It is a 5d theory like AdS/QCD.

In this letter we present
the finite temperature dynamics  in the pure gauge sector derived from the setup of \cite{ihqcd}.
We find  that  this setup describes very
well the  basic features of large-$N_c$ Yang Mills at finite temperature.
It exhibits  a first order deconfining phase transition.
The equation of state and
speed of sound  of the high
temperature phase are  remarkably similar to the corresponding lattice results. Moreover,
using  the zero temperature potential and without adding any extra
parameter, we obtain a value for the
 the critical temperature  in
very good agreement  with the one computed from the lattice.
A detailed derivation of the results  will appear elsewhere, \cite{GKNL2}.

\section{Improved Holographic QCD at T=0}

The holographic  model introduced in \cite{ihqcd} is five-dimensional.
The  basic fields that are non-trivial
in the vacuum solution, and describe the pure gauge dynamics,
  are the 5d metric $g_{\m\n}$,
a scalar $\Phi$ (the dilaton) that controls the 't Hooft coupling $\l_t$
of QCD, and an axion $a$, that is dual to to the QCD $\theta$ angle.
Moreover, as the kinetic term of the axion is suppressed by 1/$N_c^2$, it does not play any role neither in the
geometry, nor in the evolution of the 't Hooft coupling. It has however a non-trivial profile in the vacuum, implying an IR running
of the effective $\theta$-angle, \cite{ihqcd}.
Quarks can be added to the pure gauge theory by adding space-filling $D_4-\bar D_4$ brane pairs in the background gauge theory solution.
The $D_4-\bar D_4$ tachyon condensation then induces chiral symmetry breaking, \cite{ckp,ihqcd}.

The action for the 5D Einstein-dilaton theory reads,
  \begin{equation}
   S_5=M_p^3 N_c^2\le(-\int d^5x\sqrt{g}
\left[R-{4\over 3}{(\partial\l)^2\over \l^2}+V(\l) \right]+2\int_{\partial M}d^4x \sqrt{h}~K\ri)
  \label{action}\end{equation}
where $M_p$ is the Planck mass \footnote{The physical Planck mass that governs the interactions is $M_p N^{2\over 3}$. We will however call
$M_p$ the Planck mass for simplicity.} and we use the conventions of \cite{book}. The
second term in the action is the Gibbons-Hawking with $K$ being the extrinsic curvature on the boundary.

The only nontrivial input in the two-derivative action of the
graviton and the dilaton is the dilaton potential $V(\l)$, where
$\l=e^{\Phi}$. $\l$ is proportional to the 't Hooft coupling of
the gauge theory, $\l=\kappa \l_t$. The  constant of
proportionality $\kappa$ cannot be calculated at present from
first principles but as we discuss below all of the physical
observables turn out to be independent of $\ka$. The potential is
directly related to the gauge theory $\beta$-function once a
holographic definition of energy is chosen. Although the shape of
$V(\l)$ is not fixed without knowledge of the exact gauge theory
$\beta$-function, its UV and IR asymptotics can be determined.

In the UV, the input comes from perturbative QCD. We demand asymptotic freedom
 with logarithmic running.
This implies in particular that the asymptotic UV geometry is that of $AdS_5$ with logarithmic corrections.
It requires a (weak-coupling) expansion of $V(\l)$ of the form
$V(\l) = 12/\ell^2 (1 + v_1 \l + v_2 \l^2 +\cdots)
$.
 Here
$\ell$ is the AdS radius and $v_i$ are dimensionless parameters of the potential directly related to the perturbative $\beta$-function coefficients of QCD,
\cite{ihqcd}. In conformal coordinates, close to the  $AdS_5$ boundary at
$r=0$,  the metric and dilaton behave  as \footnote{We will use a ``zero'' subscript to indicate quantities evaluated at zero temperature.}:
\bea\label{sol0UV}
    ds^2_0 &=& \frac{\ell^2}{r^2} \le(1+\frac89\frac{1}{\log r\La}+\cdots\ri)\le(dr^2+dx_4^2\ri),\\
    \l_0 &=& -\frac{1}{\log r\La}+ \cdots\nn
\eea
where the ellipsis represent higher order corrections that arise from second and higher-order terms
in the $\beta$-function. The mass scale $\La$ is an initial condition for the dilaton equation and corresponds to
$\Lambda_{QCD}$.

Demanding confinement of the color charges restricts the large-$\l$ asymptotics
of $V(\l)$.  In  \cite{ihqcd} we focused on potentials such that, as $\l\to \infty$,
$ V(\l) \sim \l^{\frac43}(\log{\l})^{(\a-1)/\a}$
where $\a$ is a positive parameter. The  IR asymptotics of the solution in the Einstein frame are:
\begin{equation}\label{sol0IR}
    ds^2_0 \to e^{- C \le(\frac{r}{\ell}\ri)^{\a}}\!\!\le(dr^2+dx_4^2\ri),
    \quad\l_0 \to e^{3C/2 \le(\frac{r}{\ell}\ri)^{\a}}\!\!\le(\frac{r}{\ell}\ri)^{\frac34(\a-1)}
\end{equation}
      where the constant $C$ is a positive constant related to $\La$ in (\ref{sol0UV}). Confinement
requires $\a \geq 1$.
 The parameter $\a$ characterizes the large excitation
      asymptotics of the glueball spectrum, $m_n^2\sim n^{2(\a-1)/\a}$. For linear confinement, we choose $\a=2$.

The parameters of the holographic model a priori are: the Planck mass $M_p$,  which governs the scale of interactions between the glueballs in the theory,
 $\ka$ that relates $\l$ and the 't Hooft coupling, the parameters $v_i$ that specify the shape of the potential, the scale $\La$
that plays the role of $\Lambda_{QCD}$ and the AdS scale $\ell$.
The latter  is not a physical parameter but only a choice of
scale: only $\La\ell$ enters into the computation of physical
observables. A specific choice  for  $V(\l)$ was made in
\cite{ihqcd} with the appropriate asymptotic properties, that only
depended on a single parameter which can be taken as $v_1$, hence
fixing all $v_i$ for $i>1$. Furthermore, one can show that all
of the physical observables both at zero T and finite T are left
invariant under a rescaling of $\l$. More concretely, given a
potential $V(\l)$ and a dilaton profile that follows from this
potential with an integration constant $\La$, there exists another
profile with a different integration constant $\La_{\eta}$ which
follows from a rescaled potential $V_{\eta}(\l) = V(\eta\l)$ and
the two solutions yield the same glueball spectra and the same
thermodynamic observables. This symmetry allows one to scale away
the parameter $\ka$. Finally, $v_1$ and $\La$ are fixed by
matching to the lattice data for the first two $0^{++}$ glueball
masses. Once $\Lambda$ is fixed, all other interesting scales,
like the fundamental string scale $\ell_s$ and the effective QCD
string tension $\sigma$ are also fixed.

This determines all the parameters of the theory except the Planck mass $M_p\ell$.
We shall show below that $M_p$ can be indirectly inferred from the large  temperature behavior.

\section{The deconfinement transition}

At finite temperature
there exist two distinct types of solutions to the action (\ref{action}) with AdS asymptotics, (\ref{sol0UV}):
\begin{enumerate}
  \item[i.] The thermal graviton gas, obtained by compactifying the Euclidean time in the zero temperature solution with  $\tau\sim \tau+1/T$ :
\begin{equation}\label{TG}
    ds^2 = b^2_0(r)\le(dr^2 + d\tau^2+ dx^2_3\ri), \,\, \l=\l_0(r).
\end{equation}
This solution exists for all $T\geq 0$ and corresponds to a confined phase,
if the gauge theory at zero T confines.

\item[ii.] The black hole (BH) solutions (in Euclidean time) of the form:
\begin{equation}\label{BH}
    ds^2 = b^2(r)\le(\frac{dr^2}{f(r)} + f(r) d\tau^2+ dx^2_3\ri), \,\, \l=\l(r).
\end{equation}
The function $f(r)$ approaches unity close to the boundary at $r=0$. There exists a singularity in the interior
at $r=\infty$ that is now  hidden by a regular horizon at $r=r_h$ where $f$ vanishes.  Such  solutions correspond to a  deconfined phase.

\end{enumerate}
 As we discuss below, in confining theories
the  BH solutions exist only above a certain minimum temperature, $T>T_{min}$.


The thermal gas solution has two parameters: T and $\La$.
The black hole solution should also have a similar set of parameters:
 the equations of motion are
second order for $\l$ and $f$, and first order for $b$ \cite{GKNL2}.
Thus, {\em a priori} there are 5 integration constants to be specified. A combination of two integration constants of $b$ and $\l$
determines $\La$. (The other combination can be removed by reparametrization invariance in $r$). The condition $f\to 1$ on the
boundary removes one integration constant and demanding regularity at the horizon, $r=r_h$, in the form $f\to f_h(r_h-r)$, removes
another. The remaining integration constant can be taken as $f_h$,
related to the temperature by $4\pi T= f_h$.
From Einstein's equations one can show \cite{GKNL2}:
\begin{equation}\label{T}
4\pi\,T= b^{-3}(r_h)\le(\int_{0}^{r_h} {du\over b(u)^3}\ri)^{-1}.
\end{equation}

In the large $N_c$ limit, the saddle point of the action is dominated by one of the two types of solutions.
In order to determine the one with minimum free energy, we
need to compare the actions evaluated on solutions i. and ii. with equal
temperature.

We introduce
a cutoff boundary at $r/\ell=\eps$ in order to regulate the infinite volume.
The difference of the two scale factors is given near the boundary as
\cite{GKNL2}:
\begin{equation}\label{bb0}
    b(\eps) - b_0(\eps) = \mathcal{C}(T) \eps^3+\cdots
\end{equation}
By the standard rules of AdS/CFT we can  relate $\mathcal{C}(T)$ to the difference of VEVs of the gluon condensate:
$\mathcal{C}(T) \propto\la\tr F^2\ra_T - \la\tr F^2\ra_0 $.

The free energy difference is given by \cite{GKNL2}:
\bea
{\cal F} &=& M_p^3N_c^2V_3\le(15 \mathcal{C}(T)\ell^{-1} -\pi T b^3(r_h)\ri)\nn\\
{} & = & 15 \mathcal{C}(T)\,M_p^3N_c^2V_3\ell^{-1} -{T S \over 4},\label{free2}
\eea
where, in the last equality, we used the fact that the entropy is given by the area of the horizon.
It is clear that the existence of a non-trivial deconfinement phase transition is driven by a non-zero value for the thermal gluon condensate $\mathcal{C}(T)$.

For a general potential we can prove the following statements, that only
require the validity of the laws of black hole thermodynamics:

\begin{itemize} \item[i.] {\em There exists a
phase transition at finite T, if and only
if the zero-T theory confines.}  \item[ii.] {\em This transition is
of the {\bf first order} for {\bf all} of the confining geometries,
with a single exception described in iii:}
\item[iii.] {\em In the limit
confining geometry $b_0(r)\to \exp(-C r)$ (as $r\to \infty$), the phase
transition is of the {\bf second order} and happens at $T =
3C/4\pi$.}
\item[iv.] {\em All of the non-confining geometries at zero T are always in the
black hole phase at finite T. They exhibit a second order phase
transition at $T=0^+$.}
\end{itemize}

We now sketch a  heuristic argument, limited to
asymptotics of the type   (\ref{sol0IR}). A general, coordinate
independent  proof will appear in \cite{GKNL2}.

The existence
of  a minimum black hole temperature  $T_{min}$  in confining theories follows from the small and large $r_h$ behavior of the geometries.
On one hand, the black-hole
approaches an AdS-Schwarzschild geometry near the boundary,
 which obeys $T=1/\pi r_h$.
On the other hand, as the horizon approaches the deep interior \ie
$r_h\to \infty$, the mass of the black-hole vanishes and the black
hole solution approaches the zero-$T$ geometry in this limit. In
passing, we note that  this  implies vanishing of ${\cal F}$ in
this limit. Using the large $r_h$ limit in (\ref{T}), we find the
following asymptotics for $T$: \be\lab{Tbigrh} T\to
\frac{3C\a}{4\pi} r_h^{\a-1}, \; r_h\to\infty ; \quad T\to
\frac{1}{\pi r_h}, \; r_h\to 0. \ee

The large $r_h$ behavior in eq. (\ref{Tbigrh}) is valid under
the assumption that the zero-$T$ solution, with IR asymptotics
 (\ref{sol0IR}),  can be continuously deformed
into a black hole with arbitrarily small mass and arbitrarily
large value of $r_h$. This assumption indeed holds, as we will show elsewhere  \cite{GKNL2}
for a more general class
of confining backgrounds.

\begin{figure}
 \begin{center}
\includegraphics[height=6cm,width=9cm]{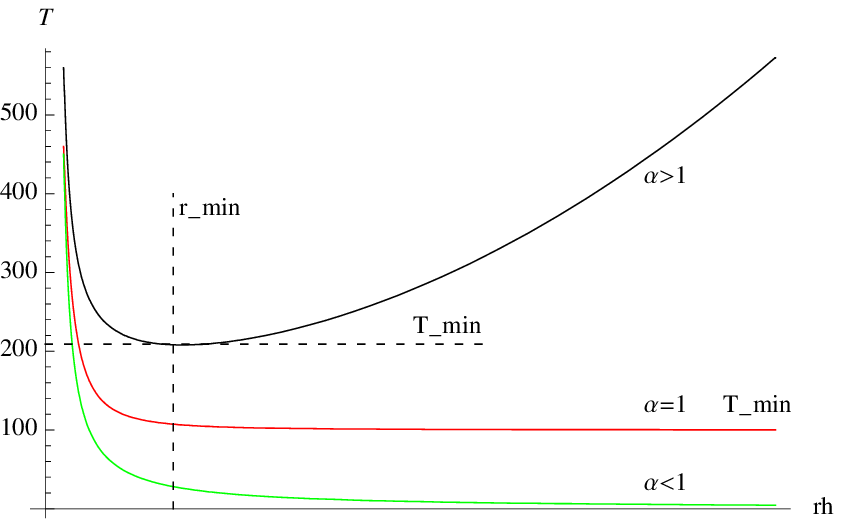} \hspace{1.3cm}(a)\\
\vspace{0.8cm}
\includegraphics[height=6cm,width=9cm]{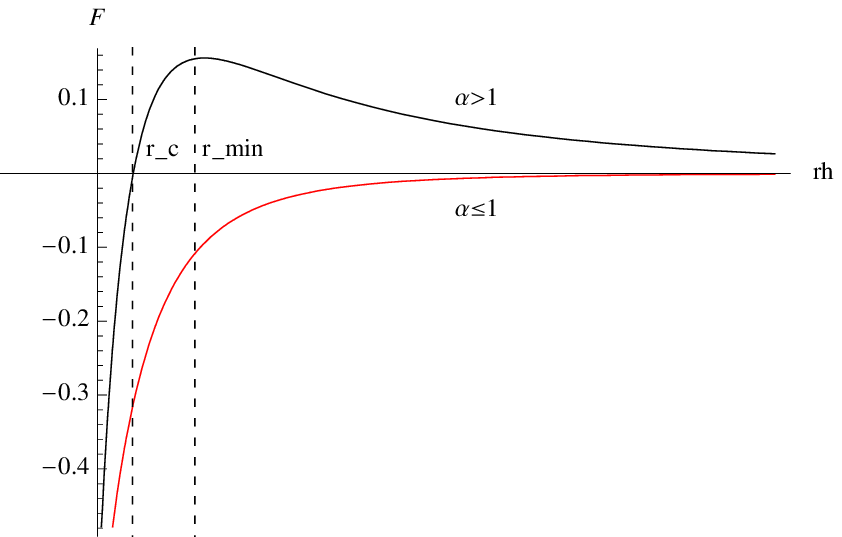} \hspace{1.3cm}(b)
 \end{center}
 \caption[]{Schematic behavior of temperature (a) and  free energy density (b)
as a function of $r_h$, for the infinite-$r$ geometries of the
 type (\ref{sol0IR}), for different values of $\a$.}
\label{illus}
 \end{figure}

Eq. (\ref{Tbigrh}) shows that for
$\a\geq 1$, that there exists a minimum temperature $T_{min}>0$
 above which the black-hole solutions exist.
Here, for
simplicity,  we assume
a single extremum of the function $T(r_h)$.
We
illustrate the function $T(r_h)$ schematically in figure
\ref{illus}.
The simple convex shapes in (a) are due to our assumption of a single minimum. In general
the function $T(r_h)$ may exhibit multiple extrema. Our demonstration here can be generalized to these
cases \cite{GKNL2}.
In the
confining geometries $\a>1$, for a given $T>T_{min}$, there  exist a big and a
 small black hole solution, given by
$r_h<r_{min}$ and $r_h>r_{min}$ respectively, see fig.\ref{illus}. The big BH has
positive specific heat hence it is thermodynamically stable, whereas the small BH is unstable.
In the borderline confining geometry $\a=1$, there is a single BH solution.

Existence of a  critical temperature $T_c\geq T_{min}$  for $\a
\geq 1$ follows from the physical requirement of positive entropy.
From the first law of thermodynamics, it follows that $d{\cal
F}/dr_h = -S\, dT/dr_h$. Then, as $S>0$ for any physical system,
extrema of ${\cal F}(r_h)$ should coincide with the extrema of
$T(r_h)$. Using also the fact that ${\cal F}(r_h)\to -\infty$ for
$r_h\to 0$ and ${\cal F}(r_h)\to 0$ near $r_h\to \infty$, we
arrive at conclusion (ii) described above: {\em There is a first
order transition for all of the confining geometries}.

An interesting case is the borderline confining geometry, where
$T_c$ coincides with $T_{min}$ and located at $r_h=\infty$. The
entropy vanishes there because the geometry shrinks to zero size.
The free energy also vanishes because this point coincides with
$T_c$. Therefore the latent heat also vanishes and one has a {\em
second order transition}. Although this geometry is not
interesting for the gauge theory, it is of some interest for GR.
We recall \cite{ihqcd}, that it corresponds to an asymptotically
AdS geometry that becomes a linear dilaton background in the deep
interior. We have shown that such a geometry exhibits a second
order Hawking-Page transition into a black-hole solution. By
similar arguments, point iv of the proposition above can also be
demonstrated without difficulty.

Finally, the small $r_h$ asymptotics  also allows us to fix the
value of the Planck mass in (\ref{action}). Small $r_h$
corresponds to high $T$. This geometry corresponds to an ideal gas
of gluons with a free energy density ${\cal F}\to (\pi^2/45) N_c^2
V_3 T^4.$ On the other hand, as the geometry becomes AdS,  eq.
(\ref{free2}) implies\footnote{It can be shown that the first term
in (\ref{free2}) is subleading in the high T limit.} that: ${\cal
F}\to \pi^4 (M_p\ell)^3 N_c^2 T^4 V_3.$ Hence we conclude that,
\begin{equation}\label{planck}
    M_p\ell = \le(45\pi^2\ri)^{-\frac13}.
\end{equation}
Using the value of $\ell$ in \cite{ihqcd}, we obtain $M_p\approx
2.32\,$ GeV.

\section{Numerical Results}

In \cite{ihqcd} an explicit form of the scalar potential with the correct asymptotics  was proposed.
The resulting background, that corresponds to the choice $\a=2$ in (\ref{sol0IR}), exhibits asymptotic freedom,  linear confinement, and
a glueball spectrum in very good quantitative agreement with the lattice
data.
Here we present a numerical computation of the relevant thermodynamic quantities
in the same theory.

The potential chosen in \cite{ihqcd} was fixed such that the UV expansion reproduces
the Yang-Mills beta-function up to two loops and has the large-$\l$ asymptotics
$V(\l) \sim \l^{4/3}(\log\l)^{1/2}$. It depends on two parameters:
the first is the  overall normalization (that fixes the
$AdS$ length $\ell$ and the energy units); the second is  $b_0$, that is
equivalent to the coefficient of linear term in the small $\l$ expansion, i.e. $v_1$.
These parameters were fit
to reproduce the lattice results for the two lowest scalar glueball masses.

Our general analysis shows that this theory has black hole solutions
above a temperature $T_{min}$ and exhibits a first order phase transition
at some $T_c>T_{min}$

To analyze the behavior of the theory at finite temperature, we have solved numerically
Einstein's equations for the metric and dilaton. The integration constants were fixed
as explained earlier.
We find a  minimum temperature for the existence of black hole solutions, $T_{min}=210$ MeV.

\begin{figure}[h]
 \begin{center}
\includegraphics[scale=0.7]{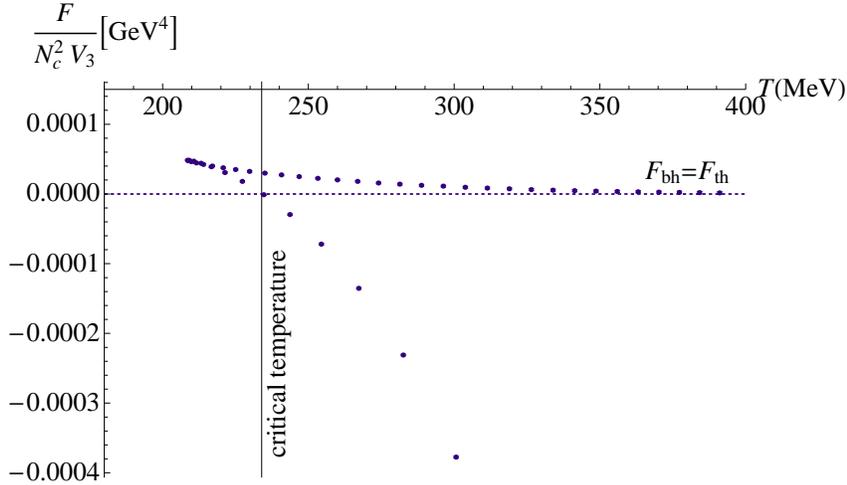}
 \end{center}
 \caption[]{Black hole free energy}
\label{FT}\end{figure}

Next, we compute the free energy difference between the black hole and thermal
gas solutions, as a function of temperature. As shown in eq. (\ref{free2}), there
are two competing contributions, which must be dealt with separately:
\begin{enumerate}
\item The term $\pi T b^3(r_h)$ can be obtained directly by evaluating the  numerical solution
at the horizon.
\item The term  $15 \mathcal{C}(T)\ell^{-1}$  must be
extracted by fitting the coefficient of the cubic   term in  the black hole
scale  factor close to the boundary, $b(r) - b_0(r) \sim \mathcal{C}(T) r^3$.
This is a large source of error in our numerics, since it is a tiny quantity
arising as a difference of $O(1)$ quantities.
\end{enumerate}
The resulting free energy as a function of the temperature is shown in figure \ref{FT}, which
clearly shows the existence of a minimum temperature, and  a first order
phase transition at $T=T_c$, where ${\cal F}(T_c) = 0$. For $T<T_c$,
the thermal gas dominates, and the system is in the confined phase. For $T>T_c$, the
(large) black hole dominates, corresponding to a deconfined phase. The
 small black hole branch is thermodynamically disfavored at all temperatures.

The value we obtain for the critical temperature,  $T_c = \bm{235\pm 15}$ MeV, is close to
the value obtained for large-N Yang-Mills \cite{teperlucini},
which with our normalization of the lightest glueball  would be $260\pm 11$ MeV \footnote{The physical units are obtained by fixing $m_{0++}=$1475\ MeV as
in \cite{ihqcd}. The value $260\pm11$ MeV is obtained combining the
results in \cite{teperlucini} and \cite{teperlucini2}}.
It should be emphasized that, we did not have to adjust any new parameter with
respect to the zero-temperature theory in order to obtain this result.

 From the free energy we  can determine all other quantities by thermodynamic identities.
However, for numerical precision it is preferable to derive the entropy directly as
the black hole area,  rather than as a derivative of
 the free energy. The latter suffers from the uncertainty in the determination of ${\cal C}(T)$. Also, due to the linear
 dependence of all thermodynamic quantities on $V_3$, it is
convenient to use densities. The pressure, and the energy and entropy densities of the
deconfined phase are given by:
\be\label{densities}
p  = -{\cal F}/V_3, \quad  s = 4 \pi M^3_p N_c^2 b^3_T(r_h), \quad \epsilon = p + T s .
\ee

 \begin{figure}[h]
 \begin{center}
\includegraphics[scale=0.7]{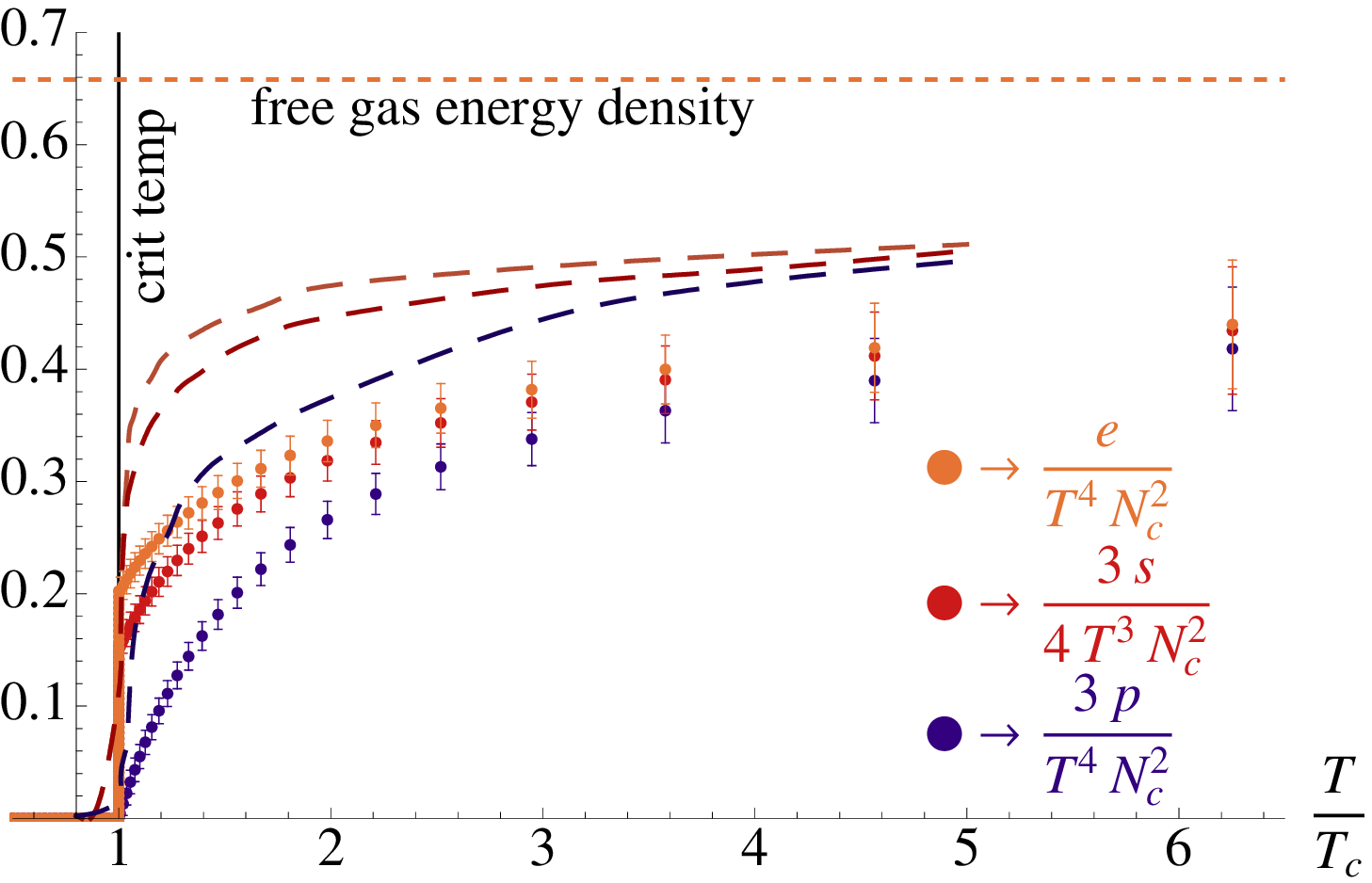} \hspace{0.5cm} (a) \\
\includegraphics[scale=0.7]{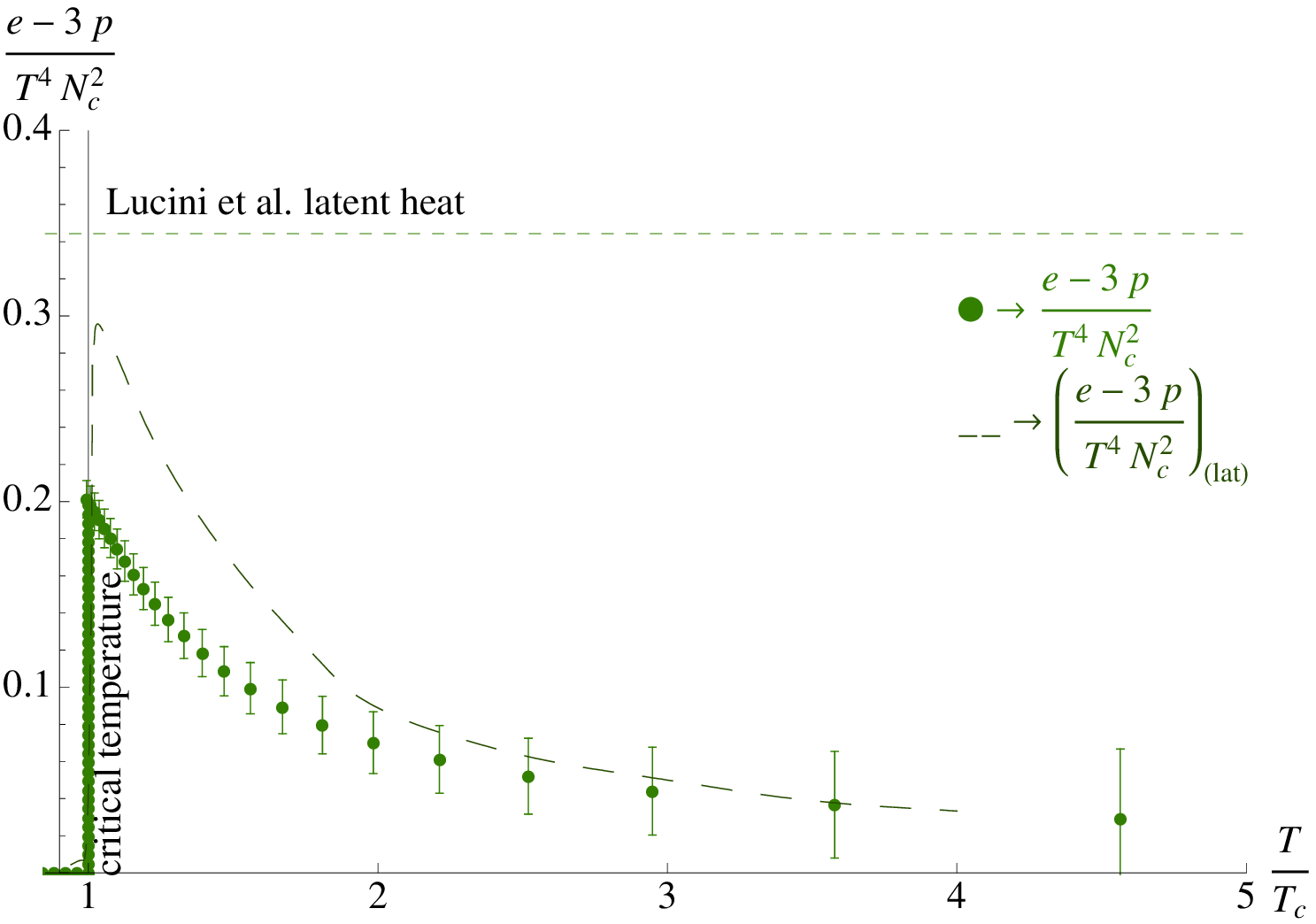} \hspace{0.5cm} (b)
 \end{center}
 \caption[]{(a) Dimensionless thermodynamic functions and
 (b)  interaction measure. The dashed curves correspond to the lattice
 data of \cite{karsch}}
\label{eps}\end{figure}

Next, we present  some of the thermodynamic quantities that are  compared with
the lattice results. It is useful to compare dimensionless quantities,
so that the $\ell$-dependence drops out.

{\bf Latent Heat}
The latent heat per unit volume is defined as the jump
in the energy at the phase transition,  $L_h = T_c\Delta s(T_c)$, and it is expected
to scale as $N_c^2$ in the large $N_c$ limit \cite{teperlucini}.
From eq. (\ref{densities}) we note that this expectation is reproduced
in our theory. Quantitatively, we find $L_h^{1/4} /T_c \simeq 0.65 \sqrt{N_c}$.
This is to be compared with the value $0.77$ reported in \cite{teperlucini}.

{\bf Equation of state and the interaction measure.}
A useful indication about the thermodynamics of a system is given by the relations
between the quantities $\epsilon/T^4$, $3 (p/T^4)$, $3/4(s/T^3)$ (the normalizations
are chosen so that they all equal the same constant in the case of a free relativistic gas).
In figure \ref{eps} (a) we compare our results for these quantities with the corresponding
lattice results, reported in \cite{karsch}\footnote{These results are for $N_c=3$; we
are unaware of similar plots obtained in the large $N_c$ limit.}.
In the low temperature phase, the thermodynamic functions vanish to the leading order in $N_c^2$
and the jump in $\epsilon$ and $s$ at $T_c$ reflects the first order phase transition.

The  {\em interaction measure}, $(\epsilon-3p)/T^4$ 
(proportional to the trace anomaly), is plotted
in figure \ref{eps} (b), together with the lattice result from \cite{karsch}.
From eq. (\ref{free2}), $\epsilon - 3p \propto \mathcal{C}(T)$, consistent
with our interpretation of ${\cal C} (T)$ as the gluon condensate.

\begin{figure}[h]
 \begin{center}
\includegraphics[scale=0.7]{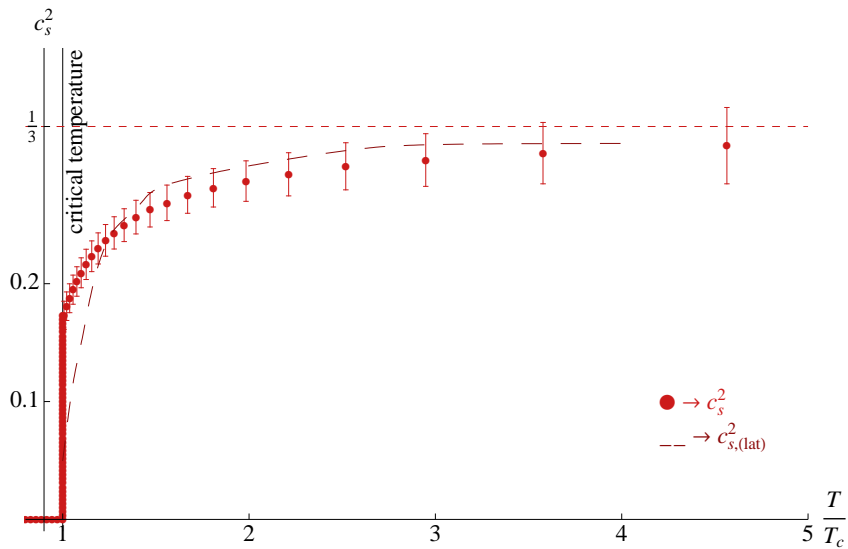}
 \end{center}
 \caption[]{Comparison between the speed of sound in our model and the lattice result of \cite{karsch} (dashed curves)}
\label{cs}\end{figure}

{\bf Speed of sound.}
This quantity  is defined as $c_s^2 = (\partial p / \partial \epsilon)_{S} = s/c_v$. It is
expected to be small at the phase transition, and to reach the conformal value $c_s^2 = 1/3$ at high temperatures.
In figure \ref{cs} we compare our results with the lattice data, finding
good agreement.

{\bf Shear viscosity.} In agreement with the general results of
\cite{Buchel}, the ratio between shear viscosity and entropy density
is $\eta/s = (4\pi)^{-1}$.

\section{Discussion}
The model presented here describes
well the  basic features of large-$N_c$ Yang Mills at finite temperature:
it exhibits  a first order deconfining phase transition, and the
temperature dependence
of the pressure, entropy, energy  density, interaction measure and
speed of sound in the high
temperature phase  behave similarly to the corresponding
lattice results.
Without adding any extra parameter, one obtains a value for
 the critical temperature  10 \% off the lattice value.

On the other hand the model  can be improved in many ways. The
latent heat $L_h/T_c^4$ is 40\% off the lattice value. Also,
our comparison
shows that (see e.g. fig 3a) approach to the free field limit at high T is slower than the lattice data.
This may be traced back to the relative smallness of the latent heat in our potential.
Although the UV and the IR asymptotics of the dilaton potential
are fixed by general requirements from the field theory, the intermediate region is free to modify.
The reason is that the low-level glueball spectrum and the thermodynamics near the phase transition are not controlled by the same
regions  of the potential.
With a suitable deformation one hopes to obtain better agreement with the lattice data. In particular,
it is possible to obtain a fit to quantities in figs. 3 and 4, well within the errors of the lattice data
in a temperature range $T_c<T<5 T_c$ \cite{GKNL2}.
Retrofitting the potential is an interesting challenge that we plan to address in \cite{GKNL2}.


 \section*{Acknowledgments}

 We thank K. Rajagopal and M. Teper for useful discussions.
This work was partially supported  by European Union Excellence Grant,
MEXT-CT-2003-509661. U.G. and F.N.  are  supported by
European Commission Marie Curie Fellowships, contracts MEIF-CT-2006-039962
and MEIF-CT-2006-039369. L.M. is supported by INFN fellowship and has partially been supported by ICTP.

\vspace{1.5cm}
{\bf Note added}
While this paper was
being  written, the work \cite{gubser} appeared, discussing related
issues in a similar setup.

\end{document}